\documentclass[%
 reprint,
 amsmath,amssymb,
 aps,
]{revtex4-1}
\usepackage{float}
\usepackage{graphicx}% Include figure files
\usepackage{dcolumn}% Align table columns on decimal point
\usepackage{bm}% bold math
\usepackage{mathtools}
\usepackage[utf8]{inputenc}

\begin{document}

%\preprint{APS/123-QED}

\title{Unitarity flow in 2+1 dimensional massive gravity}% Force line breaks with \\

\author{Sinan Sevim}
 \email{sinan.sevim@itu.edu.tr}
\author{M. Salih Zöğ}
 \email{zog@itu.edu.tr}
\affiliation{%
 Department of Physics, Istanbul Technical University, \\
 Maslak 34469 Istanbul, Turkey 
}%

\date{\today}% It is always \today, today,

\begin{abstract}
We analyze the most general case of third-order Chern–Simons-like theories of massive 3D gravity. Results show the conditions for finding the unitary regions on the parameter space. There exists $(n-1)${th} order theories on the boundary of a unitary $n$th order model on parameter space under certain conditions, as the first example recently demonstrated from the bimetric generalization of exotic massive gravity. We investigated the mechanism that causes this type of transition for third-order models. Hamiltonian analysis of the theory also presents that ghost and no-ghost regions can be separated by Chern–Simons theories.
\end{abstract}

\maketitle

%\tableofcontents

\section{Introduction}
General relativity (GR) can be considered as a consistent theory of nonlinearly interacting massless spin-2 particles, or gravitons, on four-dimensional space-time. On the other hand, GR in three dimensions (3D) yields a trivial theory without any local degree of freedom \cite{djt}. However, a degree of freedom is attainable by adding a gravitational Chern–Simons term to an Einstein–Hilbert action in 3D, which is called topological massive gravity (TMG) \cite{d-j}. This modification generates a massive bulk mode in the linearized theory which emerges from the third-order derivative acting on the metric tensor. The new degree of freedom corresponds to helicity +2 or -2 massive graviton, depending on the overall sign factor, with positive energy and obeys causality.

As seen in TMG, the number of degrees of freedom in the theory can be increased by adding higher-order derivate terms to an Einstein–Hilbert action. Another example, new massive gravity (NMG) \cite{nmg}, is a parity-preserving fourth-order model and its linearization gives the Fierz–Pauli theory \cite{fp} with two massive graviton modes with helicity $\pm 2$. If one abandons the parity-preserving condition on NMG the theory can be extended to general massive gravity (GMG) \cite{cslike2} that has $\pm 2$ helicity states with different masses and gives both TMG and NMG as a limit. Other than NMG and GMG, there is another fourth-order theory called exotic massive gravity \cite{emg}, which is a parity-odd theory with an intriguing feature called third-way consistency \cite{3rd}. Instead of adding higher-order derivative terms, a bimetric theory can be formulated. In the 3D case, the first-order formulation of the bimetric theory is known as zwei-dreibein gravity (ZDG) \cite{zd} that has a limiting case to NMG.

A motivating reason to study 3D gravity is its profound relation with 2D conformal field theory (CFT). The relation has been shown that under certain conditions 3D gravity theory generates two copies of Virasoro algebra on the boundary dual to the bulk theory \cite{b-h}. The duality is also known as the AdS/CFT correspondence \cite{ads1,ads2}. This connection permits us to work on the boundary CFT instead of bulk space. Although, in this frame, we need to deal with a problem known as bulk/boundary clash. Even if TMG, NMG, and GMG are seen as unitary theories, their dual CFTs are nonunitary due to the negative central charges. An attempt to have a unitary CFT from these theories ends up with a nonunitary bulk theory which contains negative energy gravitons. Minimal massive gravity (MMG) \cite{mmg} solved this problem in a novel manner by adding a specific combination of higher-order curvature terms to TMG.

The nondynamical nature of GR in 3D permits it to be seen as topological theory and hence a Chern–Simons gauge theory \cite{witten,a-t}. The Chern–Simons(CS) formulation allows us to have a toy model for quantum gravity and gather valuable knowledge to get insight on 4D. The CS formulation is also a convenient tool to form higher-order theories in first-order formalism. In this case, the new theory is called “CS-like” \cite{cslike} and contains a degree of freedom instead of being a topological theory. In CS-like terminology, the theory contains first-order equations of motion equal to the number of flavors. These equations can be summed up to a single differential equation order of the number of flavors. Hence, the order of theory will be used interchangeably to refer to flavor.  In this manner, higher-order theories can be achieved by increasing the number of flavors in the theory. In the TMG case the theory can be constructed by adding an extra flavor to the Einstein–Cartan action, which only contains vielbein (e)  and dual-spin connections ($\omega$) to form a third-order or three-flavor theory.

In this work, the most general three-flavor case is analyzed. By most general, we mean a model with a Lagrangian, which includes all the possible terms with arbitrary coefficients, up to three derivatives. At this point, equations of motion are not “simply solvable” like cases \cite{nmg,mmg,emg}; one may need to use infinite serial expansions for several regions. However, generic cases have more opportunities to become a unitary model.

In our case, there are ten linearly independent terms in Lagrangian with their arbitrary coefficients, which can be seen as parameters. Since we only consider the models that contain AdS background solution, two of these coefficients become dependent via additional restrictions. Therefore, our parameter space is eight-dimensional and every theory is represented here by a point. Between two points (or models) there exist infinitely many trajectories as usual, thus on these paths, we can observe a theory that smoothly flows from one to another. During the flow, the model passes from different segments with various characteristics, including unitarity. Here, we conclude that unitary and nonunitary regions are separated by seven-dimensional hypersurfaces, which can intersect with a flow at a transition point. These hypersurfaces contain lower-order theories, as the first example recently demonstrated by \cite{bigrav}. Here, we will investigate the general properties of the $n=3$ model by means of bulk/boundary unitarity and local degrees of freedom. After the study of the theory, we are expecting to observe $3 \rightarrow 2$ transitions at critical surfaces.

\section{Chern–Simons-like Formalism}

The 3D gravity models can be expressed in terms of one form $a^r$, where in three-flavor case $r=1,2,3$, and we have $a^r=(e,f,\omega)$. Here $f$ is an auxiliary field and $e,\omega$ denotes the dreibein and dual spin connection, respectively. Defining the torsion and dual curvature two-forms,
\begin{equation}\label{tr}
T(\omega)=D(\omega)e=de+\omega \times e, \ \ R(\omega)=d\omega +\frac{1}{2} \omega\times\omega
\end{equation} by the exterior derivative. One can construct the Lagrangian three-form with a flavor metric $g_{rs}$ and a coefficient tensor $f_{rst}$ as mentioned in \cite{cslike}. \begin{equation}\label{lag}
L=\frac{1}{2} g_{r s} a^{r} \cdot d a^{s}+\frac{1}{6} f_{r s t} a^{r} \cdot\left(a^{s} \times a^{t}\right),
\end{equation} where dot and cross product notation is used for three-dimensional local frame indices.
\begin{equation}
    (a^r\times a^s)^a\equiv\epsilon^{abc}a^r_b a^s_c, \ a^r \cdot a^s \equiv \eta^{ab} a^r_{\ a} a^{s}_{\ b}.
\end{equation} Following the procedure introduced in \cite{emg}, most general models in three-flavor theory have \begin{equation}\label{n3lag}\begin{split}
L=&\frac{a_1}{2} e\cdot T+a_2e\cdot Df+a_3e\cdot R+\frac{a_4}{2} f\cdot Df+\\&a_5f\cdot R+a_6 L_{LCS}+\frac{a_7}{6} e\cdot e\times e+\frac{a_8}{2} e\cdot f\times f\\&+\frac{a_9}{2} e\cdot e\times f+\frac{a_{10}}{6} f\cdot f\times f,
\end{split}
\end{equation}
 where the Lorentz–Chern–Simon three-form defined as $ L_{LCS}=\frac{1}{2} \omega \cdot \left( d\omega + \frac{1}{3} \omega \times \omega \right)  $. One can see the components of the coefficient tensor as,for instance, $f_{122}=a_8$. Therefore, all the calculations after this step will stand on these $a_i$-values. For example, the equations of motion of (\ref{n3lag}) are,\begin{equation}\label{eom}\begin{split}
 &\delta e: a_1 T+a_2 Df+ a_3 R+\frac{a_7}{2} e \times e +\frac{a_8}{2} f\times f+a_9 e\times f=0\\
 &\delta  \omega: a_3 T+a_5 Df+ a_6 R+\frac{a_1}{2} e \times e +\frac{a_{4}}{2} f\times f+a_2 e\times f=0\\
 &\delta f: a_2 T+a_4 Df+ a_5 R+\frac{a_9}{2} e \times e +\frac{a_{10}}{2} f\times f+a_8 e\times f=0,
 \end{split}
 \end{equation}
which can be summarized with the following general form, \begin{equation}\label{er}
E_r=g_{r s} d a^{s }+\frac{1}{2} f_{r s t}a^{s} \times a^{t}.
\end{equation} Although this method seems natural, it causes the later expressions to be extremely long and unclear. Passing to a new set of coefficients can ease them all. One can raise the index of $E_r$ with the flavor metric
 \begin{equation}
E^p=g^{pr} E_r=  d a^{p }+\frac{1}{2} f^{p}_{\ s t}a^{s} \times a^{t}.
\end{equation} Notice that, now, the derivatives are alone and the coefficients are $f^{p}_{\ s t}$ instead of $f_{r s t}$. We can define a new set of coefficients to write equation of motion by using (\ref{er})

\begin{equation}\label{curv_eom}
\begin{split}
&T+\frac{1}{2}\alpha_1 e\times e+\alpha_2 e\times f+\frac{1}{2}\alpha_3 f\times f=0\\
&Df+\frac{1}{2}\alpha_4 e\times e+\alpha_5 e\times f+\frac{1}{2}\alpha_6 f\times f=0\\
&R+\frac{1}{2}\alpha_7 e\times e+\alpha_8 e\times f+\frac{1}{2}\alpha_9 f\times f=0,\\
\end{split}
\end{equation}
where $\alpha$'s are  \begin{equation}
f^{1}_{\ 11}=\alpha_1, \ \ f^{1}_{\ 12}=\alpha_2, \ \
f^{1}_{\ 22}=\alpha_3 \ ...  \ \ etc.
\end{equation}
Thanks to these new $\alpha$ parameters, our separated equations of motion can be written in an elegant fashion. One can check necessity of the $\alpha$'s by just looking at \begin{equation}
 \alpha_1=\frac{ a_ 1 a_ 2 a_ 5-a_ 1 a_ 3 a_ 4-a_ 2 a_ 6 a_ 9+a_ 3 a_ 5 a_ 9+a_ 4 a_ 6 a_ 7-a_ 5^{2} a_ 7 }{ det(g)},
\end{equation}
where $det(g)$ is the determinant of flavor metric.

\subsection{Linearization about AdS}
In AdS background, our auxiliary flavor $f_{\mu\nu}$ has to be proportional to the dreibein, therefore, \begin{equation}
e=\bar{e}, \ \ f=\bar{f}=c \bar{e}, \ \ \omega=\bar{\omega},
\end{equation} where $c$ is the proportionality constant and determined by the “AdS conditions”, which we will investigate in a general sense. According to above definitions, one can write the first-order fluctuations as \begin{equation}
e=\bar{e}+\kappa k, \ \ f=c(\bar{e}+\kappa k)+ \kappa p , \ \ \omega=\bar{\omega}+\kappa v .
\end{equation}
With the help of $\alpha_i$ coefficients, linearized equations of motion from (\ref{curv_eom}) is derived:
\begin{equation}\label{lineq}\begin{split}
&\bar{D}k+\bar{e}\times v+(\alpha_3c+\alpha_2)\bar{e}\times p=0\\
&\bar{D}p+M \bar{e}\times p=0\\
&\bar{D}v+\frac{1}{l^2} \bar{e}\times k + (\alpha_9c+\alpha_8)\bar{e}\times p=0,
\end{split}
\end{equation} where $M=\alpha_5+(\alpha_6-\alpha_2)c-\alpha_3c^2$ is the mass parameter.This can be tested through some of the well-known 3D models, such as TMG \cite{d-j} and MMG \cite{mmg}. In TMG case, it can be directly seen that (\ref{lineq}) gives the correct mass parameter. However, MMG needs to redefine the spin connection $\Omega=\omega+\alpha f$ as in \cite{mmg}.

One can notice that in terms of $a_i$-coefficients, our mass parameter will be horrific. In addition to being simple, the main advantage of $\alpha_i$ formalism is its prediction of $M$'s denominator. Since all $\alpha_i$ have flavor metric determinant on the denominator, then $M$'s denominator must be proportional to $det(g)$. This feature is crucial for the analysis of unitarity, as will be shown later.

By using the expressions of torsion and curvature two-forms in the AdS background, 
\begin{equation}
\bar{T} \equiv D(\bar{\omega}) \bar{e}=0, \quad \bar{R} \equiv R(\bar{\omega})=\frac{1}{2} \Lambda \bar{e} \times \bar{e},
\end{equation}
where $\Lambda$ is the cosmological constant. The following relations which we refer to here as AdS conditions can be derived from equations of motion:
\begin{equation}\label{ads}
\begin{aligned}
&\frac{1}{2}\alpha_1+\alpha_2 c+\frac{1}{2} \alpha_3 c^2=0\\& \frac{1}{2}\alpha_4+\alpha_5 c+\frac{1}{2}\alpha_6 c^2=0,\\& \frac{1}{2}\alpha_7+\alpha_8 c+\frac{1}{2} \alpha_9c^2=\frac{1}{2l^2}.
\end{aligned}
\end{equation}
\subsection{Unitarity conditions}
In this section we will examine the bulk and boundary unitarity of the most general three-flavor model.  Our conditions are the positivity of central charges (if one considers the Brown–Henneaux condition for $AdS_3$), absence of tachyons and ghost degrees of freedom.

For boundary, we have to first obtain the central charges. By the method introduced in \cite{cc} one can find,\begin{equation}
C_\pm=\frac{3l}{2G} (-(a_3+a_5 c)\pm\frac{1}{l}a_6).
\end{equation} Therefore, $C_\pm>0$ is the sufficient condition for boundary unitarity.

For bulk, one has to find the Fierz–Pauli mass $\mathcal{M}$ which is related to the mass parameter we found in (\ref{lineq}) as
$\mathcal{M}^2 =M^2+\Lambda$. After this observation, no-tachyon condition $\mathcal{M}^2>0$ becomes $M^2l^2>1$, which now can be connected to the $a_i$ coefficients. Absence of ghost degrees of freedom needs a little more effort as it has to diagonalize (\ref{lineq}) and its Lagrangian. By the redefinitions,\begin{equation}
\begin{split}
&k=l(f_+-f_-)+\frac{l^{2}(M \alpha_ 3 c+M \alpha_ 2+\alpha_ 9 c+\alpha_ 8)}{l^{2} M^{2}-1} p\\
&v=f_+ + f_-+\frac{\alpha_9 c l^{2} M+\alpha_ 8 l^{2} M+\alpha_ 3 c+\alpha_ 2}{l^{2} M^{2}-1} p,\\
\end{split}
\end{equation} (\ref{lineq}) can be written in a more simple form, \begin{equation}
\begin{split}
&\bar{D}f_++\frac{1}{l}\bar{e}\times f_+=0\\
&\bar{D}f_--\frac{1}{l}\bar{e}\times f_-=0\\
&\bar{D}p+M \bar{e}\times p=0,
\end{split}
\end{equation} which also admits that the second-order perturbative Lagrangian becomes also diagonal. \begin{equation}\label{diaglag}
\begin{split}
L^{(2)}=&l a_+ f_+ \cdot (\bar{D}f_++\frac{1}{l}\bar{e}\times f_+)-l a_- f_- \cdot (\bar{D}f_--\frac{1}{l}\bar{e}\times f_-)\\
&-\frac{K}{M} p\cdot (\bar{D}p+M \bar{e}\times p),
\end{split}
\end{equation} where $a_{\mp}= - \frac{2G}{3l} C_\pm$ and $K=\frac{M det(g)}{2 a_+a_-}$ are written in well-known parameters. It can be concluded that no-ghost condition is $K>0$ as one goes after the same analysis of MMG in \cite{mmg}.

\subsection{Hamiltonian analysis}
The third-order theories can have either 0 or 1 degree of freedom, which can be counted by analyzing their primary and secondary constraints. In this section we will follow the method introduced in \cite{hamilton}. To begin with, we must choose a timelike hypersurface to decompose the Lagrangian to time and space components.  In this case one-forms and the Lagrangian decompose as \begin{equation}\begin{split}\label{decomp} & a^{r a}=a_{0}^{r a} d t+a_{i}^{r a} d x^{i}\\& L=-\frac{1}{2} \varepsilon^{i j} g_{r s} a_{i}^{r} \cdot \dot{a}_{j}^{s}+a_{0}^{r} \cdot \phi_{r},
\end{split}\end{equation}
where $\phi_r$ and $a_0^r$ are the constraints and  Lagrange multipliers of the system, respectively, and $\dot{a}^{s}_{i}$ is the differentiated with respect to the time component, 
\begin{equation}\phi_{r}^{a}=\varepsilon^{i j}\left(g_{r s} \partial_{i} a_{j}^{s a}+\frac{1}{2} f_{r s t}\left(a_{i}^{s} \times a_{j}^{t}\right)^{a}\right).\end{equation}
According to (\ref{decomp}), one can define the Hamiltonian and postulate the Poisson brackets of the system, \begin{equation}\begin{split} 
&\mathcal{H}=-\frac{1}{2} {e}^{i j} g_{r s} a_{i}^{r} \cdot \dot{a_{j}}^{s}-\mathcal{L}=-a_{0}^{r} \cdot \phi_{r}\\
&\left\{a_{i a}^{r}(x), a_{j b}^{s}(y)\right\}_{\mathrm{P.B.}}=\varepsilon_{i j} g^{r s} \eta_{a b} \delta^{(2)}(x-y).\end{split}\end{equation} To get further on the constraint analysis, we must  define the “smeared” test functional by integrating constraints with a test function $\xi$ on a spacelike hypersurface $\Sigma$,
\begin{equation}\phi[\xi]=\int_{\Sigma} d^{2} x \xi_{a}^{r}(x) \phi_{r}^{a}(x).\end{equation}
However , it must be checked to see if the smeared operators have well-defined functional derivatives at the boundary. In this case, improved operator $\psi[\xi]$ can be defined by adding a boundary counter-term $Q[\xi]$ to eliminate the boundary terms on the variation,
\begin{equation}\varphi[\xi]=\int_{\Sigma} d^{2} x \xi_{a}^{r} \phi_{r}^{a}+Q[\xi].
\end{equation}
 After these adjustments, finally we can calculate the Poisson brackets of improved primary constraint operators of the theory \begin{equation}\begin{split}\label{pb_const}
\{\varphi[\xi], \varphi[\eta]\}=& \phi[[\xi, \eta]]+\int_{\Sigma} d^{2} x \xi_{a}^{r} \eta_{b}^{s} \mathcal{P}_{r s}^{a b} \\
&-\int_{\partial \Sigma} d x^{i} \xi^{r} \cdot\left(g_{r s} \partial_{i} \eta^{s}+f_{r s t} a_{i}^{s} \times \eta^{t}\right),
\end{split}\end{equation}
 where $[\xi, \eta]^{t}=f_{r s}^{\ \ t} \xi^{r} \times \eta^{s}$. The first integral of the above expression contains a crucial object $\mathcal{P}_{r s}^{a b}$ named as the primary Poisson bracket matrix which has the form
 \begin{equation}\label{pmat} \mathcal{P}_{r s}^{a b}=f_{q\left[rs\right] p t}^{t} \eta^{a b} \Delta^{p q}+2 f_{r[s}^{t} f_{q] p t}\left(V^{a b}\right)^{p q},
\end{equation}
 where the objects $V_{a b}^{p q}$ and $\Delta^{p q}$ have the definitions
\begin{equation}
{V_{a b}^{p q}=\varepsilon^{i j} a_{i a}^{p} a_{j b}^{q}, \quad \Delta^{p q}=\varepsilon^{i j} a_{i}^{p} \cdot a_{j}^{q}}.
\end{equation}
 Here, $\Delta^{ee}$ and $\Delta^{ff}$ are identically zero and the remaining $\Delta^{ef}$ can be investigated through the integrability conditions.

To drive the integrability conditions, we have to take the covariant exterior derivative of (\ref{eom}) and insert the below identities: \begin{equation}
D(\omega)T(\omega)=R(\omega)\times e,\ \ \ D(\omega)R(\omega)=0.
\end{equation} Then one can show that these conditions and the invertibility of dreibein impose $\Delta^{ef}=0$, which turns out to be an additional or a secondary constraint,\begin{equation}
\psi=\Delta^{ef}=0 ,
\end{equation} and has the Poisson brackets with primary constraints as \begin{equation}\begin{split}\label{poisson}
&\left\{\phi[\xi], \psi\right\}_{\mathrm{P.B.}}=P_0\epsilon^{ij}(
\xi^{f} D_{i}e_{j}-\xi^{e} D_{i}f_{j}\\& +(\alpha_4 \xi^{e}+\alpha_5 \xi^{f})e_i\times e_j +((-\alpha_1+\alpha_5)\xi^{e}+\\
&(-\alpha_2+\alpha_6)\xi^{f})e_i\times f_j +(-\alpha_2 \xi^{e}+\alpha_3 \xi^{f})f_i\times f_j),
\end{split}
\end{equation}where $P_0$ is an $a_i$ dependent coefficient (which is assumed nonzero). Calculation of (\ref{pmat}), now follows as \begin{equation}
\left(\mathcal{P}_{a b}\right)_{r s}=P_0\left(\begin{array}{cccc}{\left(Q_{a b}\right)_{r s}} & {0} \\ {0} & {0} \end{array}\right)=P_0\left(\begin{array}{cccc}{V_{a b}^{f f}} & {V_{a b}^{e f}} & {0}  \\ {V_{a b}^{f e}} & {V_{a b}^{e e}} & {0}  \\ {0} & {0} & {0}\end{array}\right).
\end{equation} The rank of matrix $\mathcal{P}_{a b}$ ( is also the same with the rank of $Q_{a b}$ ) is an important quantity to count degree of freedom. Notice that $Q_{a b}$ is $6\times 6$ and has the rank of 2 by a $Mathematica$ calculation, but it does not stand for all the analysis. To complete it, we have to also consider the Poisson brackets between primary and secondary constraints.  Now we are extending our Q-matrix to $7\times7$ with these extra relations and looking for rank again.  Taking (\ref{poisson}) as a $6\times 1$ vector and adding it and its traspose results as \begin{equation}
\mathbb{Q}=\left(\begin{array}{cc}{Q} & {v} \\ {-v^{T}} & {0}\end{array}\right), \quad v=\left(\begin{array}{c}{\left\{\phi_{e}, \psi\right\}_{P B}} \\ {\left\{\phi_{f}, \psi\right\}_{P B}} \end{array}\right)
\end{equation} The rank of $\mathbb{Q}$ is now four, unless the rare possibility of satisfying simultaneously the conditions below, \begin{equation}\label{rank}
\alpha_4=\alpha_3=0, \ \ \alpha_1=2\alpha_5, \ \ \alpha_6=2\alpha_2
\end{equation} In this situation, we have 4 secondclass and $10-4=6$ first-class constraints. Dimension of physical phase space can be counted as $3*6-6*2-4=2$. Then we can conclude that our degree of freedom is one. On the other hand, if (\ref{rank}) is satisfied, the rank of $\mathbb{Q}$ reduces to 2 and now there are 2 second-class, $10-2=8$ first-class constraints. Therefore, phase space dimension is $3*6-8*2-2=0$ and the theory propagates no bulk degree of freedom.

Notice that together with AdS (\ref{ads}) and rank-2  (\ref{rank}) , conditions impose our mass parameter $M$ to be zero. Since $M$ and $K$ are proportional to each other, one can think that as mass changes sign, no-ghost condition $K>0$ will break down. It can be concluded that Chern–Simon theories may exist on the boundaries which separate the ghost and no-ghost regions on the parameter space.

\section{Unitarity Flow}
Our main interest in this paper is the flows between unitary and nonunitary regions on the parameter space. This space is eight-dimensional, because there are ten $a_i$-parameters and three AdS conditions restrict 3 of them but regains 1. Therefore, one can conclude that the hypersurface, which separates the unitary and nonunitary regions are seven-dimensional. As first discovered in \cite{bigrav}, this surface may contain an $(n-1)$-flavor model. Here, we will investigate the generic fall of third-order theories to second ones.

First off all, we need to define what reducing to a lower order is, in a mathematical way. For instance, consider the GMG Lagrangian \cite{cslike2},
\begin{equation}
\begin{aligned}
L_{\mathrm{GMG}}=&-\sigma e \cdot R+\frac{\Lambda_{0}}{6} e \cdot e \times e+h \cdot D e\\&-\frac{1}{m^{2}} f \cdot\left(R+\frac{1}{2} e \times f\right)+\frac{1}{\mu} L_{LCS}.
\end{aligned}
\end{equation} Notice that as $m \rightarrow \ \infty$, the terms containing the auxiliary field $f_{\mu\nu}$ decouples and the theory falls down to three-flavor.Vanishing the terms' coefficients like this one is one way but not the most general one. For example, one can provide a fall by adding terms: \begin{equation}\label{dgmg}
\begin{split}
L'_{\mathrm{GMG}}=&-\sigma e \cdot R+\frac{\Lambda_{0}}{6} e \cdot e \times e+h \cdot D e\\&-\frac{1}{m^{2}} f \cdot\left(R+\frac{1}{2} e \times f\right)+\frac{1}{\mu} L_{LCS}+ f \cdot De\\& -\frac{1}{ m^2} h \cdot\left(R+\frac{1}{2} e \times h + e \times f \right).
\end{split}
\end{equation} This is a modification of GMG for no purpose. One can write the same Lagrangian with defining a new field $q_{\mu\nu}=h_{\mu\nu}+ f_{\mu\nu}$:\begin{equation}\label{rgmg}\begin{aligned}
L'_{\mathrm{GMG}}=&-\sigma e \cdot R+\frac{\Lambda_{0}}{6} e \cdot e \times e+q \cdot De\\& -\frac{1}{m^{2}}  q\cdot\left(R+\frac{1}{2} e \times q\right)+\frac{1}{\mu} L_{LCS}
\end{aligned}
\end{equation} Therefore, (\ref{dgmg}) is not actually a fourth-order model despite the fact that it has four different fields.

During a flow, coefficients of terms of a Lagrangian smoothly change and at some critical surfaces, theories fall to lower-order ones, and we get situations like (\ref{rgmg}). The main reason for this comes from the flavor metric, which is first mentioned in (\ref{lag}). As the columns of this tensor become linearly dependent on each other, different fields can be expressed in terms of the others. We know determinant $det(g)$ goes to zero where the metric degenerates, so it can be our condition for a fall:\begin{equation}
det(g)=0.
\end{equation} At these critical surfaces where the determinant vanishes, the theory can degenerate into a $(n-1)$-flavor model or a $(n-2)$ one, or totally collapse since there is no such first-order theory in 3D gravity.

Now we are expecting some of the unitarity conditions to break down while $det(g)$ goes to zero, so where the flow and hypersurface intersect, we can observe the fall. Since the mass parameter has the determinant in its denominator, it changes sign while $det(g)$ passes through zero value but $M^2$ does not. Therefore, the no-tachyon condition $M^2l^2>1$ is not a good candidate. The positivity of central charges does not depend on $det(g)$ since they are not proportional. To maintain the proportionality by hand, one has to look after the determinant after imposing $C_+=0$ or $C_-=0$. The remaining non-zero part must be zero under some certain conditions to ensure the proportionality. These conditions can be found as
\begin{equation}
\begin{split}
&a_5=+l(a_4 c+a2) \rightarrow \ det(g) \propto C_+ \\& a_5=-l(a_4 c+a2) \rightarrow \ det(g) \propto C_-.
\end{split}
\end{equation}  By the way, where we lose the positivity of one of the central charges, the determinant also vanishes. However, these types of degenerations completely collapse the theory, which is not exactly what we want.

Now we expect that the break down of the remaining no-ghost condition $K>0$ can provide a $3\rightarrow 2$ flavor fall at the transition point. Since $K=\frac{M det(g)}{2 a_+a_-}$, and mass parameter has $det(g)$ in its denominator, this condition also seems independent of metric determinant. But in practice, we observe some of the flows depend on $det(g)$ thus there must be some regions where dependence is regained. Actually, these are the ones where some factors of $K$ are proportional to the determinant. Therefore, numerator or the denominator of $K$ can have the same roots, thus the same critical points with $det(g)$.

At this point let us discuss the model when metric degenerates. Whenever this determinant goes to zero, one can understand that at least one of the metric tensor eigenvalues vanishes, thus diagonalization of $g_{rs}$ states some of the flavors will become auxiliary, because of the general form of the Lagrangian (\ref{lag}). Therefore, some equations of motion become a constraint equation with the most general form of potential terms:  \begin{equation}\label{feom}
\frac{1}{2} \alpha  e\times e+\beta e\times f+\frac{1}{2}\gamma f\times f=0.
\end{equation} Now we have to expand $e,f$ to an infinite series around the AdS background in order to find $f_{\mu\nu}$, because we have the $f\times f$ term in (\ref{feom}),
\begin{equation}
\begin{split}
&e=\bar{e}+\kappa k+\frac{\kappa^2}{2!}k'+\frac{\kappa^3}{3!}k''+\dots=\bar{e}+\sum_{n=1}^{\infty} \frac{\kappa^{n}}{n!} k^{(n-1)}\\
&f=c\left( \bar{e}+\kappa k+\frac{\kappa^2}{2!}k'+\frac{\kappa^3}{3!}k''+\dots \right)+\kappa p+\frac{\kappa^2}{2!}p'+\dots\\
&=c \left( \bar{e}+\sum_{n=1}^{\infty} \frac{\kappa^{n}}{n!} k^{(n-1)} \right) + \left( \sum_{n=1}^{\infty} \frac{\kappa^{n}}{n!} p^{(n-1)}  \right) 
\end{split}
\end{equation} where primes do not denote derivatives but higher-order fluctuations. Substituting these series into (\ref{feom}) one can show that\begin{equation}
p_{\mu\nu}=p'_{\mu\nu}=p''_{\mu\nu}=\dots =0
\end{equation} therefore, our auxiliary field becomes proportional to the dreibein; in other words, it just imposes a constraint (\ref{feom}). \begin{equation}
f=c \left( \bar{e}+\sum_{n=1}^{\infty} \frac{\kappa^{n}}{n!} k^{(n-1)} \right) = c e
\end{equation} This result matches with the fact that as $det(g) \rightarrow \ 0$, the flavor number decreases. When the metric degenerates, flavors become linearly dependent of each other.

Let us examine this result through an example. We will write two random sets of $a_i$-parameters as the first set corresponds to a unitary model and the other one is nonunitary. Of course, these sets are two points at the eight-dimensional parameter space and can be connected via infinitely many trajectories. As coefficients flow through one of these paths, $c$ also changes smoothly and its value obtained by the AdS conditions. These conditions can be expressed with $a_i$-coefficients as \begin{equation}\label{ads2}\begin{split}
&a_1+a_4 c^2+2a_2 c=\frac{a_6}{l^2}\\& a_7+a_8 c^2+ 2 a_9 c=\frac{a_3}{l^2}\\&a_9+a_{10} c^2+2 a_8 c=\frac{a_5}{l^2}
\end{split}
\end{equation} Here, we will solve $c$ from the first equation of (\ref{ads2}) by choice, and the remaining two will be used to find $a_7,a_9$. Since these two can be calculated completely by the others, parameter space dimension reduces to 8 from 10.

Notice that (\ref{ads2}) is a list of quadratic equations, which make $c$ complex at some regions. Seeing that some of the important quantities such as central charges and mass parameter depend on $c$, for simplicity, we set $a_4=0$ through the flow in this example. Thus,\begin{equation}
c=\frac{a_6-a_1 l^2}{2 a_2 l^2} \ \in {\rm I\!R}.
\end{equation} 

Now the denominator of $M$ becomes $a_2^3 \ det(g)$ and proportional to the flavor metric determinant as it claimed before. As a result, $a_2$ becomes a factor of $K$ and if we yield $a_1 a_5=0$ (at transition point), then $a_2$ is also a factor of $det(g)$.

Our random point in unitary region of parameter space has the coordinates\begin{equation}
\begin{split}
&a^{\star}_1=6, \ \ a^{\star}_2=6,  \ \ a^{\star}_3=-12, \ \ a^{\star}_4=0, \\
&a^{\star}_5=12, \ \ a^{\star}_6=-16,  \ \ a^{\star}_8=-20, \ \ a^{\star}_{10}=13 
\end{split} 
\end{equation} in AdS units $(l=1)$. The 3D gravity model with the below coordinates has positive central charges and does not include tachyons but it has a ghost degree of freedom, thus it is nonunitary: \begin{equation}
\begin{split}
&a^{\star\star}_1=22, \ \ a^{\star\star}_2=-20,  \ \ a^{\star\star}_3=-0, \ \ a^{\star\star}_4=0, \\
&a^{\star\star}_5=-40, \ \ a^{\star\star}_6=-23,  \ \ a^{\star\star}_8=13, \ \ a^{\star\star}_{10}=11 .
\end{split} 
\end{equation} In our example, the trajectory that will connect these two random points will be parametrized by the flow
\begin{equation}
a_i(\lambda)=(1-\lambda^2) a^{\star}_i +\lambda a^{\star\star}_i.
\end{equation} 
So as $\lambda$ changes 0 to 1, the theory flows through the nonunitary point from the unitary one. Then we can observe the change of central charges,metric determinant, and $K$ through the path.

\begin{figure}[H]
 \centering
\includegraphics[width=0.5\textwidth]{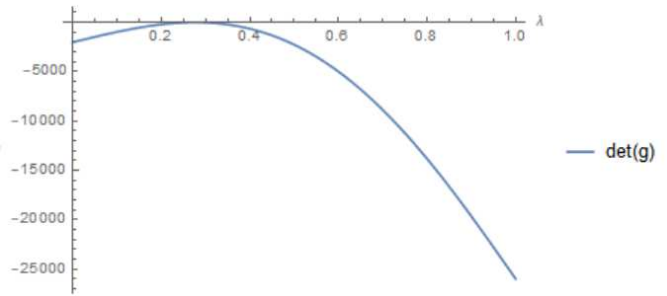}
\caption{det(g) along the flow}
\end{figure}
\begin{figure}[H]
 \centering
\includegraphics[width=0.5\textwidth]{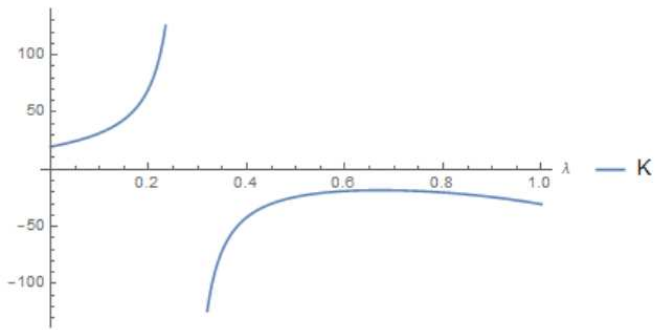}
\caption{K changes sign along the flow}
\end{figure}
\begin{figure}[H]
 \centering
\includegraphics[width=0.5\textwidth]{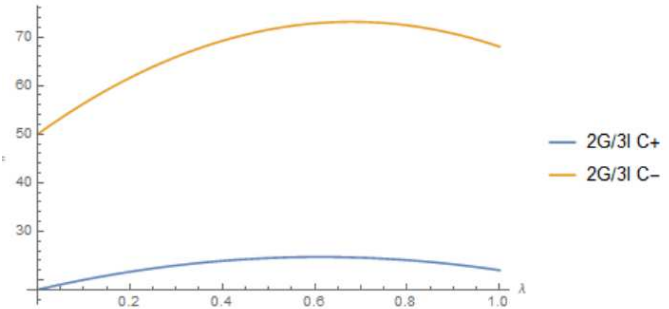}
\caption{Central charges are positive along the flow}
\end{figure}

As expected, at the point where the metric degenerates (around $\lambda \approx 0.28$), no-ghost condition $K>0$ also breaks down, and the model enters the nonunitary region. Since we know the $\lambda$ value at transition point, one can calculate the critical $a_i(\lambda)$ and find that except for $a_2,a_5,a_4$, all coefficients take nonzero values. Thus, the model's Lagrangian becomes\begin{equation}\label{rlag}\begin{split}
L=&\frac{a_1}{2} e\cdot T+a_3e\cdot R+a_6 L_{LCS}+\frac{a_7}{6} e\cdot e\times e+\frac{a_8}{2} e\cdot f\times f\\&+\frac{a_9}{2} e\cdot e\times f+\frac{a_{10}}{6} f\cdot f\times f
\end{split}
\end{equation} with equations of motion:\begin{equation}\begin{split}
&\delta e: a_1 T+ a_3 R+\frac{a_7}{2} e \times e +\frac{a_8}{2} f\times f+a_9 e\times f=0\\
&\delta  \omega: a_3 T+ a_6 R+\frac{a_1}{2} e \times e =0\\
&\delta f: \frac{a_9}{2} e \times e +\frac{a_{10}}{2} f\times f+a_8 e\times f=0.
\end{split}
\end{equation} Notice that variation according to $f$ comes in the form of (\ref{feom}); therefore, its solution gives $f=ce$, namely it becomes proportional to the dreibein. Now imposing this result into (\ref{rlag}) results in \begin{equation}\begin{split}
L=&\frac{a_1}{2} e\cdot T+a_3e\cdot R+a_6 L_{LCS}\\&+\frac{1}{6} \left( a_7+ 3 c^2 a_8 +3 c a_9+c^3 a_{10} \right) e\cdot e\times e .
\end{split}
\end{equation} Remember that $a_7,a_9$ are not free parameters, they are expressed by (\ref{ads2}). Imposing back their definitions and $\Lambda=-\frac{1}{l^2}$ now gives\begin{equation}
L=-a_6\frac{\Lambda}{2} e\cdot T+a_3e\cdot R+a_6 L_{LCS}- a_3\frac{\Lambda }{6}  e\cdot e\times e ,
\end{equation} where $a_1=\frac{a_6}{l^2}$ also by  (\ref{ads2}). Thus, equations of motion become \begin{equation}\label{neom}
\begin{split}
&\delta e:- a_6  \Lambda  T+a_3 R-a_3 \frac{\Lambda }{2}e\times e=0,\\
&\delta \omega: a_3 T+a_6 R-a_6\frac{\Lambda}{2} e\times e=0 .
\end{split}
\end{equation} Here, one can rearrange (\ref{neom}) by separating the curvatures. Finally, we get
\begin{equation}\begin{split}
&a_3 \delta e+a_6 \Lambda \delta \omega : \quad (a_3^2+a_6^2 \Lambda) \left( R-\frac{\Lambda}{2} e\times e  \right) =0\\ 
&a_3 \delta \omega - a_6 \delta e : \quad (a_3^2+a_6^2 \Lambda) \ T=0,
\end{split}
\end{equation} which areexactly the same equations as the Einstein–Cartan theory in 3D.
\section{Discussion \& Conclusion}
In this work we presented the most general form of $n=3$ Chern–Simons-like theory and derived the unitarity condition for both bulk and boundary. This classification shows us there is a hypersurface in the parameter space, which separates unitary and nonunitary regions and between them there exists a $n=2$ model under certain restrictions. One can write a flow between these regions and along this trajectory, at some transition point the theory degenerates into a lower flavor. Despite the fact that all of the study was made for $3^{rd}$ order 3D massive gravity, the ideas of degeneration and transitions are quite general and can have applications for any theory which has a matrix of its kinetic terms.

The generic case also presents that the regions which contain ghost degrees of freedom, and the regions that do not, can be separated by Chern–Simons theories. These types of models have zero mass parameters, and their constraints sufficiently construct Lie algebra. 
\newline
\paragraph*{Acknowledgements:}
We thank Mehmet Ozkan for useful discussions. The work of M.S.Z. and S.S. is partially supported by TUBITAK Grant No. 118F091.

\end{document}